\documentclass[aps,amsfonts,showpacs]{revtex4}

\usepackage{amsmath}
\usepackage{graphicx}
\usepackage[subnum]{cases}
\begin{document}

\def\be{\begin{equation}}
\def\ee{\end{equation}}
\def\bea{\begin{eqnarray}}
\def\eea{\end{eqnarray}}
\def\bml{\begin{mathletters}}
\def\eml{\end{mathletters}}
\def\b{\bullet}
\def\eqn#1{(~\ref{eq:#1}~)}
\def\no{\nonumber}
\def\av#1{{\langle  #1 \rangle}}
\def\m{{\rm{min}}}
\def\M{{\rm{max}}}

\title{Evolutionary dynamics on strongly correlated fitness landscapes}

\author{Sarada Seetharaman$^1$ and Kavita Jain$^2$}
\email{saradas@jncasr.ac.in, jain@jncasr.ac.in}
\affiliation{$^1$Theoretical Sciences Unit, Jawaharlal Nehru Centre for Advanced Scientific Research,  Jakkur P.O., Bangalore 560064, India}
\affiliation{$^2$Theoretical Sciences Unit and Evolutionary and Organismal Biology Unit, Jawaharlal Nehru Centre for Advanced Scientific Research,  Jakkur P.O., Bangalore 560064,India}

\widetext
\date{\today}

\begin{abstract}
We study the evolutionary dynamics of a maladapted population of self-replicating sequences  on strongly correlated fitness landscapes.  Each sequence is assumed to be composed of blocks of equal length and its fitness is given by a linear combination of four independent block fitnesses. A mutation affects the fitness contribution of a single block leaving the other blocks unchanged and hence inducing correlations between the parent and mutant fitness. On such strongly correlated fitness landscapes, we calculate the dynamical properties like the number of jumps in the most populated sequence  and the temporal distribution of the last jump which is shown to exhibit a inverse square dependence as in evolution on uncorrelated fitness landscapes. We also obtain exact results for the distribution of records and extremes for correlated random variables.
\end{abstract}
\maketitle

\section{Introduction}
\label{intro}

Fitness is a measure of an organism's ability to survive and reproduce \cite{Gavrilets:2004,Orr:2009}. Fit organisms produce more offspring and can dominate the population while the less fit ones can be lost. Mathematically, fitness is a non-negative real number associated with a sequence which is a string of $L$ letters whose meaning is context dependent. For example,  fitness represents the stability of a sequence of amino acids in case of proteins, activity for an enzyme or replication rate for a genetic sequence of nucleotides. On plotting the fitness as a function of the sequence, a fitness landscape is obtained. Empirical measurement of fitness landscapes is very hard since the number of sequences increases exponentially with the sequence length $L$. However several qualitative features particularly the topography of the fitness landscapes has been deduced in experiments on proteins and microbes either by an explicit construction of the fitness landscapes for small $L(\lesssim 5)$ or indirect measurements of relevant quantities. These experiments show that the fitness landscapes can have smooth hills as evidenced by fast adaptation in some proteins \cite{Romero:2009} or multiple peaks as seen in  microbial populations that evolve towards different fitness maxima \cite{Korona:1994,Burch:2000,Fernandez:2007}
 and enzymes with short uphill paths to the global fitness peak \cite{Perelson:1995}. Detailed studies in which all or a set of mutants from wild type to an optimum are created  and their fitness measured \cite{Poelwijk:2007} have also indicated the smooth \cite{Lunzer:2005} and rugged  \cite{Weinreich:2006,Visser:2009} nature of the fitness landscapes.
 
The topography of the fitness landscapes is related to the correlations between the fitness of the sequences. If the fitness of the mutants of a sequence is correlated to that of the sequence so that the mutant fitness does not differ appreciably from the parent sequence, a smooth fitness landscape is generated whereas if the mutant fitnesses are independent random variables so that the fitness of one sequence has no influence over the fitness of other sequences differing from it by even a single mutation, a highly rugged fitness landscape with multiple optima is obtained. Several theoretical models such as NK model \cite{Kauffman:1993}, block model  \cite{Perelson:1995} and rough Mt. Fuji-type model \cite{Aita:2000} in which correlations can be tuned via a parameter have been proposed. Although realistic fitness landscapes exhibit intermediate degree of correlations \cite{Carneiro:2010}, much of the theoretical work has focused on the limiting cases of fitness landscapes with high degree of correlation but single fitness peak \cite{Woodcock:1996} and no correlations but a large number of local optima \cite{Krug:2003,Jain:2005,Jain:2007c,Jain:2007a}.

In this article, we study the evolutionary dynamics on the fitness landscapes generated by the block model \cite{Perelson:1995} in which a sequence of length $L$ is assumed to be composed of independent units or blocks of length $\ell$. As explained later (see Sec.~\ref{models}), the correlations can be varied by changing the block length $\ell$ from maximally correlated case with $\ell = 1$ to maximally uncorrelated one with $\ell = L$. Here we focus on the block model with $\ell=2$ which generates fitnesses  that are strongly correlated but to a lesser degree than the maximally correlated case and  the fitness landscape is moderately rugged {\it i.e.}  exhibits several peaks.

The evolution model that we work with here describes the deterministic evolution of an infinitely large population of asexually replicating sequences. In this model, the population is   initially  distributed in such a manner that the high fitness sequences have lower initial population and vice versa but the population of all the sequences increases linearly with time  \cite{Krug:2003}. As time goes on, a highly fit subpopulation is able to overcome the poor initial condition and dominate the population until an even fitter population overtakes it. This process goes on until the globally fittest sequence becomes the most populated one. 
The stepwise dynamics of such leadership changes termed {\it jumps} have been studied when the fitness variables are completely uncorrelated \cite{Krug:2003,Jain:2005,Jain:2007c}; here we are interested in this problem when the fitnesses are strongly correlated. As explained in the next section, in the context of this problem, it is also relevant to consider the sequence with largest fitness amongst sequences carrying $D$ mutations relative to a reference sequence and  whose fitness  is a record in that its fitness exceeds the fitness of all the sequences with less than or equal to $D$ mutations. Thus we are led to study the statistics of {\it maximum}  \cite{David:2003} and {\it records} \cite{Arnold:1998} when the random variables are  not independent, both of which have been much less studied unlike the problem when the random variables are  independently distributed. 

Our detailed analysis presented in Sec.~\ref{ell2} shows that the  statistical properties studied depend only on 
whether the number of mutations $D$ are odd or even and whether $D$ lies below or above $L/2$. This simplification allows us to tackle the  problem analytically and to find exact expressions for various  quantities. On uncorrelated fitness landscapes, it has been shown that  the average number of leadership changes increases as $\sqrt{L}$ and the timing of the last jump exhibits a $1/t^{2}$ dependence \cite{Jain:2005,Jain:2007c}. For evolution on the class of strongly correlated fitness landscapes studied here, we find that the average number of jumps is a constant independent of $L$ but the time dependence of the distribution of the last jump remains unaffected. The average number of records is found to increase linearly with $L$ as in maximally rugged case albeit with a larger prefactor. 
\section{Shell model on correlated fitness landscapes}
\label{models}

Consider a microbial population evolving in a complex environment that can be modeled by rugged fitness landscapes. 
At large times, most of the population resides at the globally fittest sequence of the fitness landscape and due to mutations, a suite of mutants is also present. If the population size is infinite, a nonzero population is present at {\it all} the sequences whereas a finite population produces only a small number of mutants around the fittest sequence \cite{Jain:2007a}. Now if the environment is changed by changing (say) the nutrient medium of the microbial population, the fittest subpopulation before the environment change will be typically maladapted  to the new environment and depending on the total population size, a small population may be present at the new fittest sequence. We are interested in finding how the new global maximum is reached starting with an initial condition in which all the population is at the sequence that was globally fittest before the environmental change. The exact evolutionary dynamics of average Hamming distance and overlap
function has been studied on permutationally invariant \cite{Saakian:2008} and
uncorrelated \cite{Saakian:2009} fitness landscapes. Here we will be tracking the evolution of the {\it most populated sequence} in time on strongly correlated fitness landscapes. The dynamics of the adaptation process is studied in the setting when the population size is infinite so that the fluctuations in the population frequency of a sequence can be neglected and one can work with the averages.  
In the following, we begin with the quasispecies model of biological  evolution \cite{Eigen:1971,Jain:2007b} and proceed to relate it to the shell model  \cite{Krug:2003}. We then define and explain some properties of the block model \cite{Perelson:1995} of correlated fitness landscapes that we shall use in the paper. 

We consider an infinitely large population of binary sequences  where a sequence  $ \sigma \equiv \{\sigma_{1},...,\sigma_{L}\}~,~\sigma_i=0, 1$ is a string of $L$ letters.  The population evolves by the elementary processes of replication and mutation.  If the fitness $A(\sigma)$ of the sequence $\sigma$ is defined as the average number of copies produced per generation and $p_{\sigma\leftarrow\sigma'}$ is the probability that a sequence $\sigma'$ mutates to the sequence $\sigma$  at a Hamming distance  $D(\sigma,\sigma')=\sum_{i=1}^{L}(\sigma_{i}-\sigma'_{i})^{2}$ from it, the population fraction $X(\sigma,t)$ of sequence $\sigma$ at time $t$ evolves according to the  following quasispecies equation  \cite{Eigen:1971,Jain:2005}:
\begin{equation}
 X(\sigma,t+1)=\frac{\sum_{\sigma'}p_{\sigma\leftarrow\sigma'}A(\sigma')X(\sigma',t)}{\sum_{\sigma'}A(\sigma')X(\sigma',t)}
 \label{qs1}
\end{equation}
where the denominator on the right hand side ensures the normalisation condition $\sum_{\sigma} X(\sigma,t)=1$ is satisfied at all times. 
 Assuming that the mutations occur independently at each locus with a probability $\mu$,  the mutational probability  $p_{\sigma\leftarrow\sigma'} = \mu^{D(\sigma,\sigma')}(1-\mu)^{L-D(\sigma,\sigma')}$. 
 In the following discussion, we will use the unnormalised population $Z(\sigma,t)$ defined through the relation $X(\sigma,t)=Z(\sigma,t)/\sum_{\sigma'}Z(\sigma',t)$ as it obeys a linear equation given by
 \begin{equation}
Z(\sigma,t+1)=\sum_{\sigma'} \mu^{D(\sigma,\sigma')} (1-\mu)^{L-D(\sigma,\sigma')} A(\sigma') Z(\sigma',t)
\label{qs2}
\end{equation}

As discussed at the beginning of this section, we consider the evolution of the dominant population starting with the initial condition $X(\sigma,0)=Z(\sigma,0)=\delta_{\sigma,\sigma^{(0)}}$ where $\sigma^{(0)}$ is the fittest sequence before the change in the environment. 
 Earlier work  \cite{Krug:2003, Jain:2005, Jain:2007c} has shown that  the statistical properties of the most populated sequence in the quasispecies model are accurately described by a simplified {\it shell model}  which approximates the solution of (\ref{qs2}) by
 \begin{equation}
 Z(\sigma,t) \sim \mu^{D(\sigma,\sigma^{(0)})}A^{t}(\sigma) 
 \label{shell0}
 \end{equation}
The above equation can be heuristically obtained as follows: on iterating (\ref{qs2}) with the given initial condition, the population 
$ Z(\sigma,1) \sim \mu^{D(\sigma,\sigma^{(0)})}A(\sigma^{(0)})$ for small $\mu$. 
Thus  all the mutants become available in one generation for an infinitely large population even after starting with a highly localised population. 
If the mutations are neglected for further evolution i.e. $Z(\sigma,t+1)=A (\sigma) Z(\sigma,t)$, the solution (\ref{shell0}) is immediately obtained. 
A detailed analysis has shown that the behavior of $Z(\sigma,t)$  in shell model matches the quasispecies dynamics (\ref{qs2})  only for highly fit sequences and at short times. However it captures the behavior of the most populated genotype  {\it exactly} at all times \cite{Jain:2007c} and therefore we will work with the shell model in the rest of the article. 

Taking the logarithm of both sides of (\ref{shell0}) and rescaling the time by $|\ln \mu|$, the logarithmic population $E(\sigma,t) \sim \ln Z(\sigma,t)$ is seen to increase linearly in time with a slope $\ln A(\sigma)=w(\sigma)$, 
\begin{equation}
 E(\sigma,t)=-D(\sigma,\sigma^{(0)})+w(\sigma) t
 \label{shell1}
\end{equation}
According to the above equation, there are ${L \choose D}$ populations in a shell of radius $D$ from the initial sequence which have the same initial condition but different growth rates. As  the fittest population in each shell grows the fastest, one can work with the largest fitness $w^{(\max)}(D)$ in each shell. Labeling the fittest sequence in a shell by its shell number,  (\ref{shell1}) can be rewritten as
\begin{equation}
 E(D,t)=-D +w^{(\max)}(D) t
 \label{shell2}
\end{equation}
Thus we arrive at a model in which the fitness variables $w^{(\max)}(D)$ are independent but non-identically distributed. 
We mention in passing that the above linear dynamics when the slope variables are independent and identically distributed (i.i.d.) have appeared in a shell model with one-dimensional fitness \cite{Krug:2003}, a gas of elastically colliding hard core particles \cite{Bena:2007}  and  a spin glass model with random entropy  \cite{Krzakala:2002}.

As mentioned earlier, we are mainly interested in the dynamics of the most populated sequence  whose fitness changes abruptly or jumps in time. Due to (\ref{shell2}), the leader in shell $D'$ is overtaken by a fitter population in shell $D > D'$ at time $T(D,D')$ given by
\begin{eqnarray}
 T(D,D') = \frac{D-D'}{w^{(\max)}(D)-w^{(\max)}(D')}
\end{eqnarray}
Initially the sequence $\sigma^{(0)}$ is the leader. As the overtaking time must be positive, the population in shell $D=1$ can be a leader provided $w^{(\max)}(1) > w^{(\max)}(0)$. Similarly, the fittest sequence in shell $2$ can be the most populated sequence if $w^{(\max)}(2)= \max\{w^{(\max)}(0),w^{(\max)}(1),w^{(\max)}(2) \}$. In general, a population at Hamming distance $D>0$ has a chance of becoming a leader only if its fitness is greater than that of all the other populations at Hamming distance  $D'<D$ or in other words, the fitness in shell $D$ is a record. As noted in earlier works, it is not sufficient to be a record fitness in order that the corresponding sequence can be the dominant sequence \cite{Krug:2003,Jain:2005} and a jump occurs only when the current leader is overtaken in minimum time. Due to this constraint, not all record sequences participate in the jump process and thus the number of records is an upper bound on the number of jumps. 
  
\begin{figure}
\includegraphics[angle=0,scale=0.4]{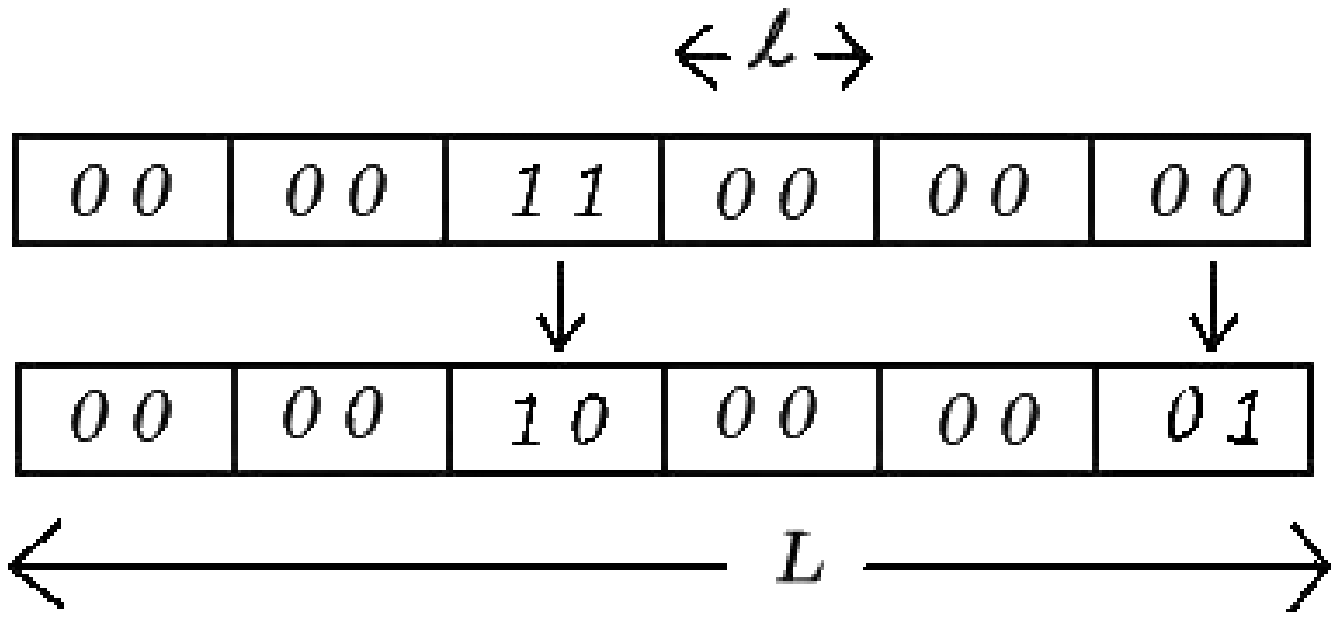}
\includegraphics[angle=0,scale=0.4]{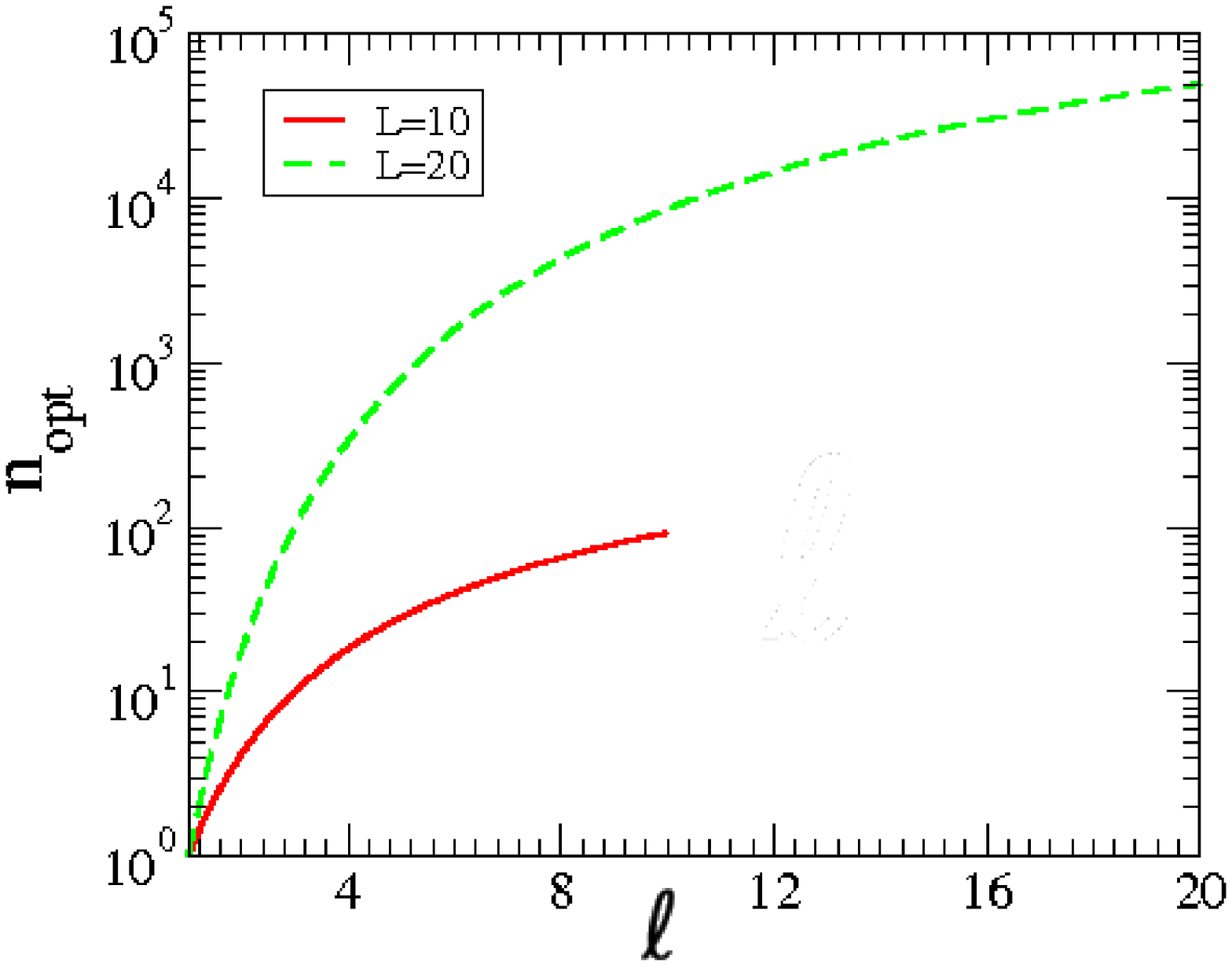}
\caption{(Color online) Left panel: Block model for $\ell=2$. The initial sequence and its mutant have correlated fitnesses as they have several blocks in common. Right panel: Average number $n_{\rm{opt}}$ of local maxima in block model as a function of $\ell$ for various $L$.}
\label{block}
\end{figure}

We next define the {\it block model} introduced by Perelson and Macken
who were motivated by the observation that many biomolecules such as
proteins and antibodies are composed of domains or partitions
\cite{Perelson:1995}.  As  shown in Fig.~\ref{block}, a sequence of
length $L$ is divided into $B$ independent blocks of equal length
$\ell=L/B~,~1 \leq \ell \leq L$. Each block configuration is assigned
a fitness value which may also depend on the position of the block
(locus-dependent block fitness model). In this article, we assume that
a block configuration at any  location in the sequence  carries the
same block fitness (see Sec.~\ref{concl} also). These $2^\ell$ block
fitnesses are chosen {\it independently} from a 
common distribution with support on the interval $\left[ l, u \right]$
where $l$ and $u$ are respectively the lower and upper limits of the
block fitness distribution. The sequence fitness  is given by the
average of the 
corresponding block fitnesses. 

The topographical features such as the number of local maxima 
depends on $\ell$. For a sequence to be a local maximum, each of its
$B$ blocks must also be a local maximum. Since a sequence is composed
of independent blocks and the average number of local optima of a
sequence of length $\ell$ with i.i.d. fitness is $2^\ell/(\ell+1)$, it
follows that the average number $n_{\rm{opt}}$ of local maxima of a
sequence of length $L$ and block length $\ell$ is given by $(
2/(\ell+1)^{1/\ell} )^L$ \cite{Perelson:1995}. Except for $\ell=1$ for
which there is a single local (same as global) fitness  peak,
$n_{\rm{opt}}$ increases with increasing $\ell$ and $L$ (see
Fig.~\ref{block}). For $\ell=2$ with which we work in this article,
there are $\approx 1.15^L$ local optima on an average. Arguing as
above for local maximum, it can be seen that  the globally fittest
sequence is composed of identical blocks with the largest block
fitness and has the average fitness given by 
$2^{\ell}\int^{u}_{l}df f p(f) [\int^{f}_{l}df'p(f')]^{2^{\ell}-1} $. Thus the initial  sequence $\sigma^{(0)}$ can be chosen to be any one of the $2^{\ell}$ sequences with same blocks. For convenience, we choose it to be the one with all $0$s.

An attractive feature of the block model is that the  correlations can be tuned with the block length $\ell$. 
As  illustrated in Fig.~\ref{block}, when two sequences have at least one common block,  their respective fitnesses are   correlated.
For $\ell=1$, the
sequence fitnesses are maximally correlated while for 
$\ell=L$, we obtain the model with maximally uncorrelated fitnesses. This statement can be quantified by considering the correlation function $C_{0,j}$ between the fitness $w_{0}=w(\sigma^{(0)})$ of the initial sequence and the fitness $w_j$  of a sequence at Hamming distance $D=1$ from it given by 
\be
w_j= \frac{(L-\ell) f_0 + \ell f_j}{L}~,~j=0,...,\ell
\ee
where $f_j$ is the fitness of the block of length $\ell$ with $1$ in the $j$th position. Using the fact  that $f_j$'s are i.i.d. variables, we can write the correlation function as \cite{Perelson:1995}
\be
C_{0,j}=\langle w_0 w_j \rangle - \langle w_0 \rangle \langle  w_j \rangle= \frac{L-\ell}{L} \sigma^2
\ee
where $\sigma^2$ is the variance of the block fitness distribution $p(f)$. The above correlation function is largest at $\ell=1$ and vanishes at $\ell=L$. Similarly 
 the correlation function $C_{i,j}$ amongst the one mutant neighbors given by \cite{Das:2010} 
\be
C_{i,j}=\langle w_i w_j \rangle - \langle w_i \rangle \langle  w_j
\rangle= \left[\left(\frac{L-\ell}{L} \right)^2+  \delta_{i,j}\left( \frac{\ell}{L}\right)^2 \right] \sigma^2
\ee
is a monotonically decreasing function of $\ell$ for $i \neq j$.

In the following, we will study the shell model defined by (\ref{shell1}) where the fitness $w(\sigma)$ is chosen from the block model. In the next section, we briefly discuss the dynamics of the shell model for the  two limiting cases namely $\ell=1$ and $L$. Section~\ref{ell2} which forms the major part of the paper discusses the evolutionary dynamics when the block length $\ell=2$. Finally we conclude with a discussion of our results in Sec.~\ref{concl}. In the rest of the article, we will assume that the sequence length $L$ is an even integer. 

\section{Shell model dynamics when block length $\ell=1$ and $\ell=L$}
\label{limits}

In this section, we briefly discuss the evolutionary
dynamics on the  fitness landscapes for the two limits of the block model namely block length $\ell=1$ and $L$. When block length is equal to one, the sequence fitnesses are maximally correlated. Let $f_0$ and $f_1$ denote the block fitness of the 
two blocks $\{ 0 \}$ and $\{ 1 \}$ respectively. Then 
the fitness $w(D)$ of a sequence at Hamming distance $D$ from
the initial sequence is given by
\be
w^{(\max)}(D)=\frac{(L-D) f_0+ D f_1}{L}
\ee
The fitness landscape thus generated is permutationally invariant
since there is a single distinct fitness at each $D$ from the initial sequence. 
It is easy to see that the average number of jumps on fitness
landscapes with $\ell=1$ is half. This is because  
if $f_{0} > f_{1}$, a  jump cannot occur after
$D=0$. If $f_{1} > f_{0}$, 
as the time taken by the population at $D > 0$ to overtake the population
at $D=0$ given by 
\be
T(D,0) = \frac{D}{w^{(\max)}(D)-w^{(\max)}(0)}= \frac{1}{f_1-f_0}~,~ D > 0
\ee
is independent of $D$, all the populations overtake $E(0,t)$ at the same time and
hence one jump occurs with probability $1/2$. Thus the average number
of jumps is $1/2$ and independent of $L$. The average number of
records from the above considerations is given by $1+(L/2)$. 

The opposite limit of maximally uncorrelated fitnesses for which $\ell=L$ has been studied earlier \cite{Krug:2003,Jain:2005,Jain:2007b}. It has been shown that the average number of records is given by $(1-\ln 2) L$ for any underlying block fitness distribution \cite{Jain:2005} and the average number of jumps by $\sqrt{L \pi}/2$ for exponentially distributed block fitnesses \cite{Jain:2007b}. 

\section{Shell model dynamics when block length $\ell=2$}
\label{ell2}

For the rest of the article, we will consider the case when the sequences are built by blocks of
length $\ell=2$. The block 
fitness is given by $f_{0}$, $f_{1}$, $f_{2}$ and $f_{3}$
corresponding to the blocks $\{0,0\},\{0,1\}, \{1,0\}$ and $\{1,1\}$ respectively. Let $n_i$ denote the
number of blocks with fitness $f_i, i=0,...,3$. Then the fitness of a sequence of length $L$ with $D$ 
mutations obtained by averaging over $B=L/2$ block fitnesses can be written as  
\bea
w_{n_{1},n_{2}}(D)=\frac{(L-D-n_{1}-n_{2})f_{0}+2n_{1}f_{1}+2n_{2}f_{2}+(D-n_{1}-n_{2})f_{3}}{L}   
\label{wdef}
\eea 
In the above expression, since the total number of
blocks equals $L/2$ and the Hamming distance $D$ of a sequence from the
initial sequence is given by $n_1+n_2+2 n_3 =D$, we get $n_3=(D-n_1-n_2)/2$
and $n_0=(L-D-n_1-n_2)/2$. As $n_0$ and $n_3$ must be integers, for
even $D$, both $n_1, n_2$ must be either even or odd whereas for odd 
$D$, either $n_1$ should be odd and $n_2$ even or vice versa. Besides,
for $D \leq L/2$, the conditions $n_1 +n_2 \leq D, n_1 \leq D$ must be
satisfied as $n_3 \geq 0$ and for $D > L/2$,  $n_1 +n_2
\leq L-D, n_1 \leq L-D$ are required to ensure the non-negativity of
$n_0$. 

As mentioned in Sec.~\ref{models}, in order to be the globally fittest sequence, a sequence must  be composed of blocks of the same type. For $\ell=2$, the global maximum can thus occur at $D=0, L/2$ and $L$ corresponding to $f_0$, either $f_1$ or $f_2$ and $f_3$  being the largest of the block fitnesses respectively. Starting with all the population at the initial  sequence, we wish to find the properties of the jumps by which the most populated sequence reaches the global maximum. In the following subsections, we discuss the statistics of extremes (Sec.~\ref{ext}), 
records (Sec.~\ref{rec}) and jumps (Sec.~\ref{jum}) on correlated fitness landscapes. 

\subsection{Distribution of the largest fitness at constant Hamming distance}
\label{ext}

It was shown in  \cite{Jain:2009} that the total number of distinct fitnesses at a fixed $D$ increases as $D^2$. However for questions of interest, we need to consider only the sequence with the largest fitness. To identify such sequences, we first consider 
fitnesses with fixed $n_1+n_2=n~,~n > 0$ where $n_1, n_2$ satisfy the
conditions described above. As the coefficient of $f_0$ and $f_3$ in
fitness $w_{n_1,n_2}$ 
depends on $n_1+n_2$, assuming $f_1 > f_2$ and comparing $w_{n_1,n-n_1}$ and
$w_{n_1',n-n_1'}$ for all $n_1' \neq n_1$,
we find that 
$w_{n_1,n-n_1} > w_{n_1', n-n_1'}$ for $n_1'  < n_1 $. 
The fitness $w_{n_1,n-n_1}$
can be the largest $w^{(\max)}_{n_1,n_2} (D) $ of all the fitnesses at fixed $D$ and $n$ provided $n_1=n$. 
 We next compare $w_{n,0}$ and  
$w_{n+k,0}~,~k \neq 0$ for $D \geq 2$. Since $w_{n+k,0}(D)=w_{n,0}(D)-(k/L) (f_0+f_3-2 f_1)$, it follows that for $D \leq L/2$, 
\begin{numcases} 
{w^{(\max)}_{n_1,n_2} (D) =}
w_{D,0}(D) & if $f_1 > f_2$ and $f_0-2 f_1+ f_3 < 0$  
\label{arg1} \\
w_{1,0}(D) & if $f_1 > f_2$ and
$f_0-2 f_1+ f_3 > 0$ and $D$ is odd 
\label{arg2} \\
w_{0,0}(D) & if $f_1 > f_2$ and 
$f_0-2 f_1+ f_3 > 0$ and $D$ is even 
\label{arg3}
\end{numcases} 
The above conditions are independent of $D$ (except for the parity) and as we shall see, this property simplifies the problem considerably. 
For $D > L/2$, the largest possible fitness is obtained on replacing
$D$ by $L-D$ in the above discussion. The corresponding conditions for the case
when $f_1 < f_2$ are  obtained by interchanging  fitnesses $f_1$ and
$f_2$ and the indices $n_1$ and $n_2$ in the preceding equation. 

\begin{figure}
\includegraphics[angle=270,scale=0.35]{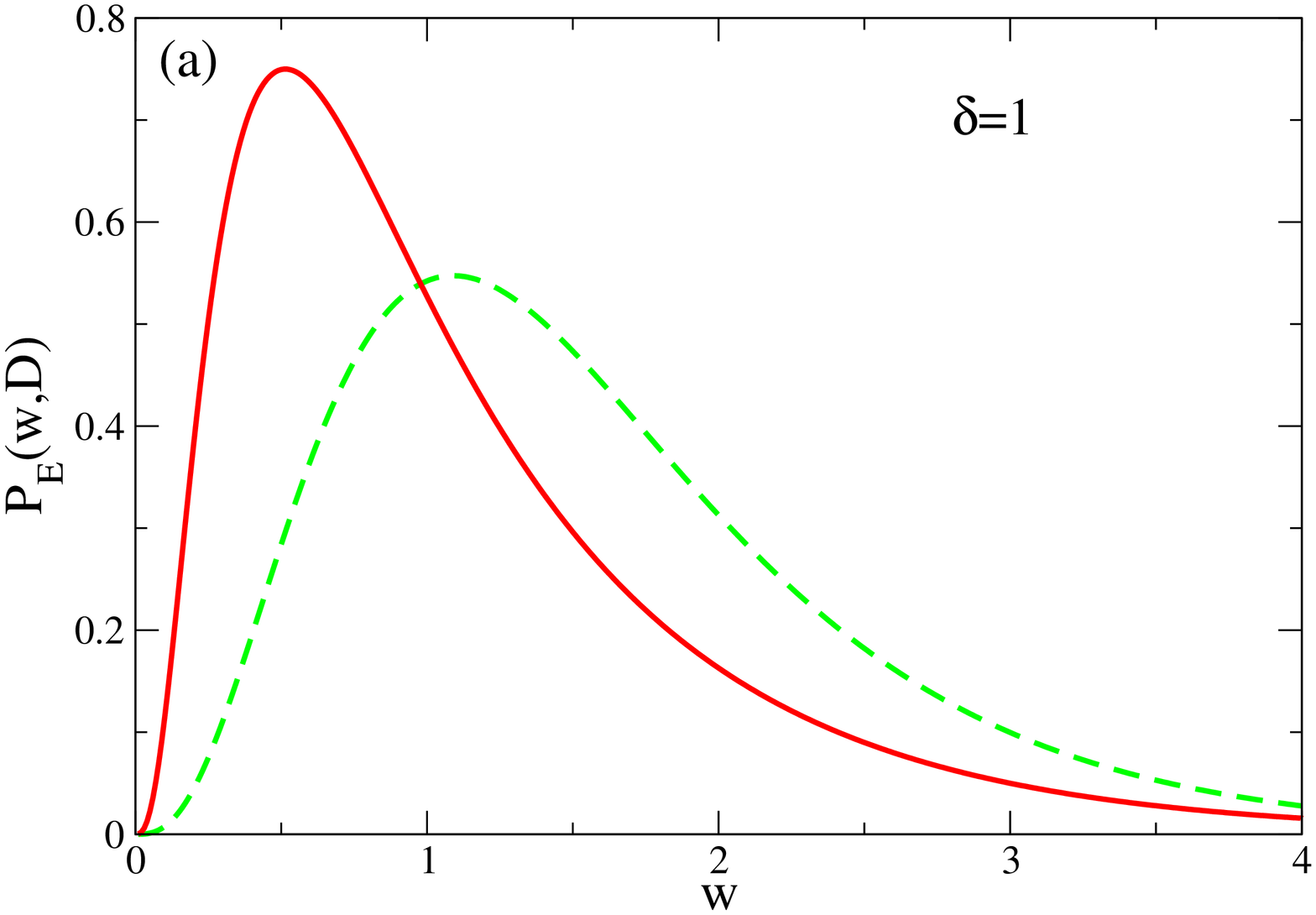}
\includegraphics[angle=270,scale=0.35]{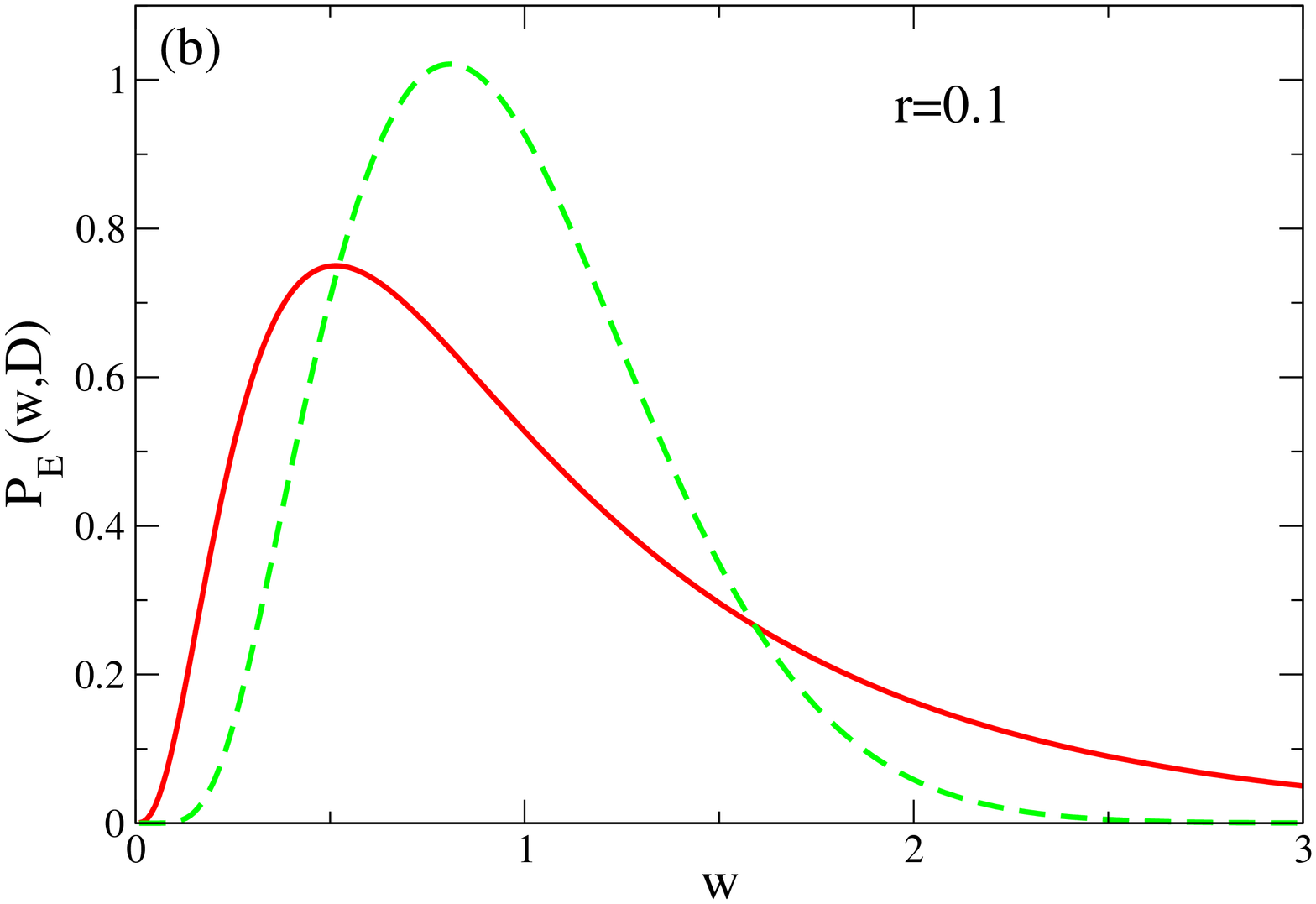}
\caption{(Color online) Distribution $P_E(w,D)$ of maximum fitness for (a) $r=0.1$ (solid) and $r=0.4$ (broken) with $\delta=1$ and (b) $\delta=1$ (solid) and $\delta=2$ (broken) with $r=0.1$.}
\label{maxfit}
\end{figure}

We consider the cumulative distribution ${\cal P}_E(w, D)$ that all the
fitnesses at constant $D$ are smaller than $w$. 
As argued above, for even $D \leq L/2$, only one of the three
fitnesses $w_{D,0}, w_{0,D}$ and 
$w_{0,0}$ can be the largest. For unbounded underlying distribution $p(f)$ with $f > 0$, we can thus write 
\bea
{\cal P}_E(w,D) &=& \int_0^\infty df_0 p(f_0) \int_0^\infty df_1 p(f_1) \int_l^u
df_2 p(f_2) \int_0^\infty df_3 p(f_3)  \Theta(w-w_{D,0})
\Theta(w-w_{0,D}) \Theta(w-w_{0,0}) \\
&=& \int_0^{\frac{w}{1-r}} df_0 ~p(f_0)
\int_0^{\frac{w-(1-r) f_0}{r}} 
df_3 ~p(f_3) \left[ \int_0^{\frac{w-(1-2 r) f_0}{2 r}} df_1 ~p(f_1) \right]^2
\eea
where $\Theta(...)$ is the Heaviside step function and $r=D/L < 1/2$. 
Specifically, for $p(f)=\delta f^{\delta-1} e^{-f^\delta},~\delta > 0$, we have
\bea
{\cal P}_E(w,D) &=& \delta \int_0^{\frac{w}{1-r}} df  f^{\delta-1} e^{-f^\delta} \left[1-e^{-
\left(\frac{w-(1-r) f}{r}\right)^\delta} \right] \left[1-e^{-\left(\frac{w-(1-2 r) f}{2 r} \right)^\delta} \right]^2 \\
&=& a \int_0^1 df  e^{-a f} \left[1-e^{-a 
\left(\frac{(1-r) (1-f^{1/\delta})}{r}\right)^\delta} \right] \left[1-e^{-a 
\left(\frac{(1-r) (1-\frac{1-2 r}{1-r} f^{1/\delta})}{2 r}\right)^\delta} \right]^2 
\label{extreme1}
\eea
where $a=(w/(1-r))^\delta$. 
The probability $P_E(w,D)=d{\cal P}_E/d w$ that the largest sequence fitness with $D$ mutations has a value $w$ can be easily computed for $\delta=1$ and is given by 
\be
 P_E(w,D)=\frac{e^{-\frac{w}{1-r}}-e^{-\frac{2 w}{r}}+2 e^{-\frac{3 w}{2 r}}}{1-2 r}-\frac{2 e^{-\frac{w}{2 r}}}{1-4 r}-\frac{\left(e^{-\frac{2 w}{1-r}}-e^{-\frac{w}{r}}\right) r}{1-5 r+6 r^2}+\frac{4 e^{-\frac{3 w}{2 (1-r)}} r}{1-6 r+8 r^2} 
\label{extvD}
\ee
The above distribution shown in Fig.~\ref{maxfit}a for two values of $r$ shifts towards right with increasing $r$ as the average $\langle w \rangle_E=1+(55 r/36)$ \cite{Jain:2009}. Figure \ref{maxfit}b shows  that the  extreme value distribution at fixed $r$ peaks at {\it larger} $w$ as $\delta$ increases. This is contrary to the general expectation that if the tail of the underlying distribution decays fast, the probability of finding a large maximum value of a set of $S$ random variables should also decrease when $S \gg 1$. Here as the number of independent random variables is merely four, the tail of the block fitness distribution is not adequately sampled and the block fitnesses lie closer to the average value which increases with increasing $\delta$ thus resulting in the behavior seen for $P_E(w,D)$. 

When $D$ is odd, one can write down an expression for the extreme distribution but for large $L$, it reduces to that obtained for even $D$. 
The results for extreme statistics when $D > L/2$ can be obtained on replacing $D$ by $L-D$ in the above discussion \cite{Jain:2009}.

\subsection{Statistics of record fitnesses}
\label{rec}

In this subsection, we are interested in finding the probability that a fitness $w_{n_1,n_2}(D)$ is a record i.e. it exceeds all the fitnesses in the shell $D' \leq D$. As only the largest 
fitness at constant $D$ can possibly be a record, we need to consider only such fitnesses.  Unless otherwise mentioned, we assume $f_1 > f_2$ so that the largest fitness at constant $D$ can be one of the following: $w_{0,0}(D)$ if $D$ is even and $w_{1,0}(D)$ otherwise, $w_{D,0}(D)$ for $D \leq L/2$ and $w_{L-D,0}(D)$ for $D > L/2$.

For $D \leq L/2$, the fitness $w_{D,0}(D)$ can be a record if it exceeds all the fitnesses at
constant $D$ as well as the ones with number of mutations $D' <
D$. The first condition is met if (\ref{arg1}) is
satisfied. As the conditions in (\ref{arg1}) are independent of $D$ (barring parity), 
the largest fitness in a 
shell with $D'$ mutations is also $w_{D',0}(D'), 1 < D' < D$. 
Then $w_{D,0}(D) > w_{D',0}(D')$ for all $D' \geq 0$ if $f_1 >
f_0$.  
Thus the probability of $w_{D,0}(D)$ being a
record can be written as 
\bea
P(w_{D,0} ~\text{is a record})&=& \int_l^{u} \prod_{i=0}^3 df_i
p(f_i)  \Theta(f_1-f_0) \Theta(f_1-f_2) 
\Theta(2 f_1-f_0-f_3)  \\
&=& \int_{l} ^{u} df_0 p(f_{0})
\int_{f_0}^{u} df_1
  p(f_{1}) \int_{l} ^{f_{1}} df_2 p(f_{2}) \int_{l}
^{2f_{1}-f_{0}} df_3 p(f_{3}) \label{recwd0}
\eea

For $D > L/2$, the fitness $w_{L-D,0}(D)$
can be record if $w_{L-D,0}(D) > w_{L-D',0}(D')$ for $D' \geq L/2$ and
$w_{L-D,0}(D) > w_{D',0}(D')$ for $D'< L/2$ along with the conditions
$f_1 > f_2$ and $f_0-2 f_1+f_3 < 0$ (see (\ref{arg1})). The first two
inequalities are satisfied if $f_3 > f_1$ and $f_0 < f_1$. Thus we can
write 
\bea
P(w_{L-D,0} ~\text{is a record}) &=& \int_l^{u} \prod_{i=0}^3 df_i
p(f_i)  \Theta(f_1-f_0) \Theta(f_1-f_2) \Theta(f_3-f_1)
\Theta(2 f_1-f_0-f_3) \\
&=& \int_l^u df_0 p(f_0) \int_{f_0}^u
df_1 p(f_1) \int_l^{f_1} df_2
p(f_2) \int_{f_1}^{2 f_1-f_0} df_3 p(f_3)  \label{recwld0}
\eea

For even $D$, the fitness $w_{0,0}(D)$ can be a
record if $w_{0,0}(D) > w_{0,0}(D')$ for even $D'$ and $w_{0,0}(D) >
w_{1,0}(D')$ for odd $D'$ besides satisfying (\ref{arg3}). If $f_2 < f_1$, 
the fitness $w_{0,0}(D)$ can be a record if $f_3 > f_0$ and $f_3 
> 2 f_1-f_0$. The last two conditions can be split into two cases, 
namely $f_3 > f_0$ if $f_0 > f_1$ and $f_3 > 2 f_1 -f_0$ if $f_0 <
f_1$. Similarly, for $f_2 > f_1$, the conditions for $w_{0,0}(D)$ to be a record are obtained by interchanging $f_2$ and $f_1$. Combining all the above conditions, we get
\bea
P(w_{0,0}~\text{is a record}) &=& 2 \int_l^{u} \prod_{i=0}^3 df_i
p(f_i)   \Theta(f_1-f_2) \Theta(f_3-f_0) 
\Theta(f_0+f_3-2 f_1)  \\
&=& 2[\int_{l}^{u} df_0 p(f_{0})
    \int_{f_0}^{u} df_1 p(f_{1}) \int_{l}^{f_{1}} df_2 p(f_{2})
    \int_{2f_{1}-f_{0}}^{u} df_3 p(f_{3}) \no \\
&+& \int_{l}^{u} df_3 p(f_{3}) \int_{l} ^{f_{3}} df_0 p(f_{0})
      \int_{l} ^{f_{0}} df_1 p(f_{1}) \int_{l} ^{f_{1}} df_2 p(f_{2})]
\label{recw00}
\eea

For odd $D$, the fitness 
 $w_{1,0}(D), D > 1$ can be a record if (\ref{arg2}) is
satisfied, $w_{1,0}(D) > w_{1,0}(D')$ for odd 
$D' < D$ and $w_{1,0}(D) > w_{0,0}(D')$ for even $D' < D$. The last
two conditions are satisfied if $f_0 < f_3$ and $f_0 < f_1$ 
respectively. Then the probability of $w_{1,0}(D), D > 1$ being a record is given by
\bea
P(w_{1,0}~\text{is a record})&=& \int_l^{u} \prod_{i=0}^3 df_i
p(f_i)  \Theta(f_1-f_0) \Theta(f_1-f_2) 
 \Theta(f_3-f_0) \Theta(f_0+f_3-2 f_1)\\
&=& \int_{l} ^{u} df_{0} p(f_{0}) \int_{f_0}
^{u} df_{1} p(f_{1}) \int_{l} ^{f_{1}}df_{2}
p(f_{2}) \int_{2f_{1}-f_{0}} ^{u}df_{3} p(f_{3})
\label{recw10}
\eea
The above expression holds for $D=1$ also as $w_{1,0}(1)$ is a record if $w_{1,0}(1) > w_{0,0}(0)$ which implies $f_0 < f_1$ besides $f_2 < f_1$.


\subsubsection{Record occurrence distribution}
\label{reco}

\begin{figure}
\begin{center}
\includegraphics[angle=270,scale=0.4]{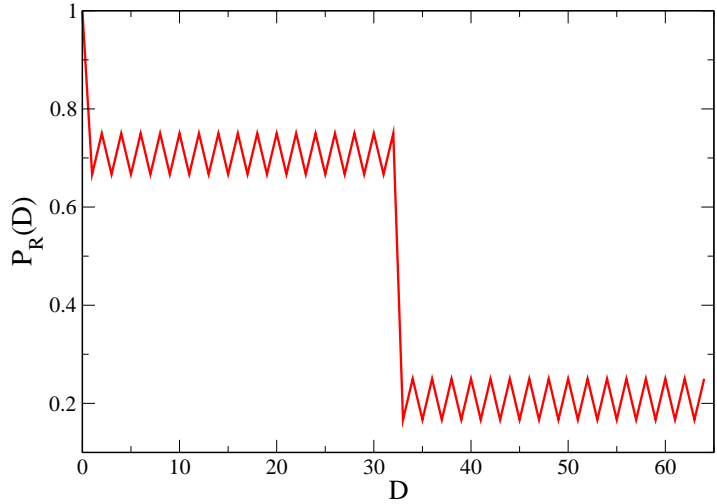}
\caption{(Color online) Variation of record occurrence probability $P_R(D)$ with Hamming distance $D$ for $L=64$.}
\label{recofig}
\end{center}
\end{figure}

Using the results derived above, we now calculate the probability $P_R(D)$ that a record occurs in the shell
with $D > 0$ mutations given $P_R(0)=1$. Figure \ref{recofig} shows that
$P_R(D)$ is not a smooth function - the value of $P_R(D)$ depends on
whether $D$ is odd or even and whether it is below or above $L/2$.  Thus four distinct cases arise due to this character of
$P_R(D)$ which we will discuss below.  We shall
find that the distribution $P_R(D)$ is universal i.e. does not  depend on the choice of the
underlying distribution of the block fitness.  As the global maximum is the last record and the only global maximum for $D > L/2$ occurs with probability $1/4$,  we may expect the record occurrence probability for $D > L/2$ to be smaller than that for $D \leq L/2$.

 \noindent \textit{Even $D$:} When $D$ is even, either $w_{D,0}(D)$ or $w_{0,D}(D)$ can be a record for $D \leq L/2$,  $w_{L-D,0}(D)$ or $w_{0,L-D}(D)$ for $D > L/2$ or $w_{0,0}(D)$ for any even $D$ .Thus the probability of even $D$ for $D \leq L/2$ having a record is given by 
\bea
P_R(D) &=& 2 P(w_{D,0}~\text{is a record} )+P(w_{0,0}~\text{is a
  record}) \\
&=&2 \int_{l}^{u} df_0 p(f_{0})
    \int_{f_0}^{u} df_1 p(f_{1}) \int_{l}^{f_{1}} df_2 p(f_{2})+ 2 \int_{l}^{u} df_3 p(f_{3}) \int_{l} ^{f_{3}} df_0 p(f_{0})
      \int_{l} ^{f_{0}} df_1 p(f_{1}) \int_{l} ^{f_{1}} df_2 p(f_{2}) \\
 &=& \frac{2}{3}+ \frac{1}{12} = \frac{3}{4} ~,~D \leq L/2
\eea

Similarly for $D > L/2$, the record occurrence probability is given by 
\bea
P_R(D) &=& 2 P(w_{L-D,0}~\text{is a record})+P(w_{0,0}~\text{is a
  record}) \\
&=&  2 \int_{l}^{u} df_0 p(f_{0})
    \int_{f_0}^{u} df_1 p(f_{1}) \int_{l}^{f_{1}} df_2 p(f_{2}) \int_{f_1}^u df_3 p(f_3)+ \frac{1}{12}
   =\frac{1}{4} ~,~D > L/2
\eea

\noindent \textit{Odd $D$:} For $w_{D,0}(D), D > 1$ to be a 
record when $D$ is odd, the same conditions as for even $D$ are required so that
(\ref{recwd0}) holds. Thus the probability 
of a shell with odd $D, 1 < D \leq L/2$ having a record is given by 
\bea
P_R(D) &=& 2\left[P(w_{D,0} ~\text{is a record})+P(w_{1,0}~\text{is a
    record})\right] \\ 
&=&  2 \int_{l} ^{u} df_{0} p(f_{0}) \int_{f_0}
^{u} df_{1} p(f_{1}) \int_{l} ^{f_{1}}df_{2}
p(f_{2}) =\frac{2}{3}~,~D \leq L/2
\eea

For $D > L/2$, the probability that 
$w_{L-D,0}(D)$ is a record is given by (\ref{recwld0}) and $w_{1,0}(D)$ is a record by (\ref{recw10}). Thus the probability of a record 
occurring for odd $D>L/2$ can be expressed as 
\bea
P_R(D) &=& 2 \left[ P(w_{L-D,0}~\text{is a record})+P(w_{1,0}~\text{is
    a record}) \right] \\
    &=&  2 \int_{l}^{u} df_0 p(f_{0})
    \int_{f_0}^{u} df_1 p(f_{1}) \int_{l}^{f_{1}} df_2 p(f_{2}) \int_{f_1}^u df_3 p(f_3)  =\frac{1}{6}~,~ D > L/2
    \label{odd2}
\eea

\subsubsection{Record value distribution}
\label{recv}

In this subsection, we calculate the probability ${\cal P}_R(w,D)$ that the record value in shell $D$ is smaller than or equal to  $w$. For this purpose, we will need the probability ${\cal P}_R(w(D) \leq w)$ that the fitness $w(D)$ in shell $D$ does not exceed $w$. As the record value distribution is not expected to be universal, we will restrict ourselves to distributions with support on the interval $[0, \infty)$. 
It can be checked that the cumulative distribution ${\cal P}_R(w,D)$ gives the probability $P_R(w)$ obtained in the last subsection when $w \to \infty$. Below we present the expressions for $D \leq L/2$ as the corresponding distributions for $D > L/2$  can be written in an analogous manner. 

\noindent \textit{Even $D$:} As seen for the distribution of extreme values in Sec.~\ref{ext}, the distribution for the record value is a function of the ratio $r=D/L$ for even $D$.  Since either $w_{D,0}(D)$ or $w_{0,0}(D)$ can be a record for even $D \leq L/2$,  the cumulative probability ${\cal P}_R(w,D)=2 {\cal P}(w_{D,0} \leq w) +{\cal P}(w_{0,0} \leq w) $ where 
 \bea
{\cal P}(w_{D,0} \leq w) &=& \int_0^{\infty} \prod_{i=0}^3 df_i
p(f_i) ~\Theta(w-w_{D,0}) \Theta(f_1-f_2) 
\Theta(2 f_1-f_0-f_3) \Theta(f_1-f_0) \no \\
&=& \int_0^w df_0 p(f_0) \int_{f_0}^{\frac{w-f_0}{2 r}+f_0} df_1
p(f_1)  \int_0^{f_1} df_2 p(f_2) \int_0^{2 f_1-f_0} df_3 p(f_3)
\label{recvalwD0}
\eea
and 
\bea
{\cal P}(w_{0,0} \leq w) &=& 2 \int_0^{\infty} \prod_{i=0}^3 df_i
p(f_i)   ~\Theta(w-w_{0,0}) \Theta(f_3-f_0) \Theta(f_1-f_2) 
\Theta(f_0+f_3-2 f_1) \no \\
&=& 2 \int_0^w df_0 p(f_0) \int_0^{f_0} df_1
p(f_1)  \int_0^{f_1} df_2 p(f_2) \int_{f_0}^{\frac{w-f_0}{r}+f_0} df_3 p(f_3) \no \\
&+& 2 \int_0^w df_0 p(f_0) \int_{f_0}^{\frac{w-f_0}{2 r}+f_0} df_1
p(f_1)  \int_0^{f_1} df_2 p(f_2) \int_{2 f_1-f_0}^{\frac{w-f_0}{r}+f_0} df_3 p(f_3) 
\label{recvalw00}
\eea
Using these expressions, it is straightforward to see that 
\bea
{\cal P}_R(w,D) &=& 2 \int_0^w df_0 p(f_0) \int_0^{f_0} df_1
p(f_1)  \int_0^{f_1} df_2 p(f_2) \int_{f_0}^{\frac{w-f_0}{r}+f_0} df_3 p(f_3) \no  \\
&+& 2 \int_0^w df_0 p(f_0) \int_{f_0}^{\frac{w-f_0}{2 r}+f_0} df_1
p(f_1)  \int_0^{f_1} df_2 p(f_2) \int_{0}^{\frac{w-f_0}{r}+f_0} df_3 p(f_3) 
\eea
Taking the derivative of the last expression with respect to $w$, we obtain the distribution $P_R(w,D)$ that the record value equals $w$. For $p(f)=e^{-f}$, the distribution $P_R(w,D)$ is given by  
\be
P_R(w,D)=\frac{e^{-4 w}+2 e^{-\frac{3 w}{2 r}}-e^{-\frac{2 w}{r}}}{1-2 r}-\frac{2 e^{-\frac{w}{2 r}}}{1-4 r}+\frac{e^{-2 w} (3-8 r)}{1-6 r+8 r^2}+\frac{e^{-\frac{w}{r}} r-e^{-3 w} (3-8 r)}{1-r (5-6 r)}
\label{recvD}
\ee

\begin{figure}
\begin{center}
\includegraphics[angle=270,scale=0.4]{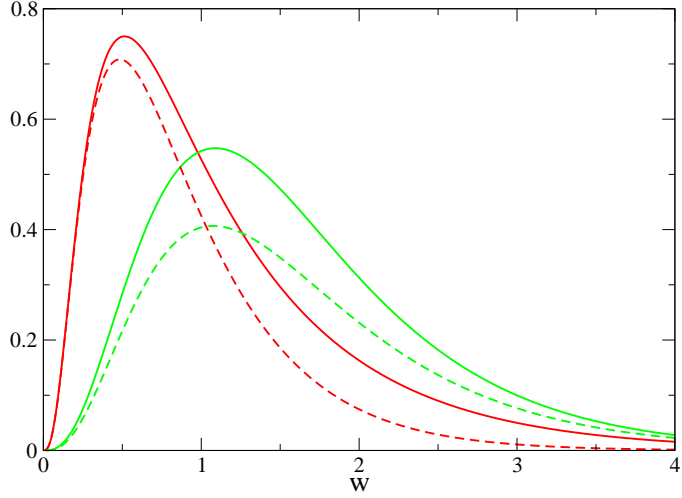}
\caption{(Color online) The probability distribution of the extreme value (solid lines) given by (\ref{extvD}) and record value (dashed lines) by (\ref{recvD}) for $r=0.1$ (left curves)  and $r=0.4$ (right curves) for $p(f)=e^{-f}$. }
\label{recvfig}
\end{center}
\end{figure}

The above result for the record value distribution is compared with the extreme value distribution $P_E(w,D)$ given by (\ref{extvD}) in Fig.~\ref{recvfig} for two values of $r$.  Though the record fitness is also the extreme fitness in shell $D$, the converse is not true and the distribution $P_R(w,D) <  P_E(w,D)$ for all $w$ at a given $D$. We also note that the most probable  record value in shell $D$ is smaller than the corresponding extreme value - this behavior is unlike that for uncorrelated fitnesses for which record is a maximum of a larger set of independent variables. 

\noindent \textit{Odd $D$:}  To find the record value distribution for odd $D$, besides ${\cal P}(w_{D,0} \leq w)$, we require the cumulative probability ${\cal P}(w_{1,0} \leq w)$ that the fitness $w_{1,0}(D)$ in shell $D$ does not exceed $w$. The latter can written as  
\bea
{\cal P}(w_{1,0} \leq w) &=& \int_0^{\infty} \prod_{i=0}^3 df_i p(f_i)~ \Theta(w-w_{1,0}) \Theta(f_1-f_2) 
\Theta(f_0+f_3-2 f_1) \Theta(f_1-f_0) \Theta(f_3-f_0) \no \\
&=& \int_0^w df_0 p(f_0) \int_{f_0}^{\frac{L(w-f_0)}{2 D}+f_0} df_1
p(f_1) \int_0^{f_1} df_2 p(f_2)
 \int_{2 f_1-f_0}^{\frac{L w-(L-D-1)f_0-2 f_1}{D-1}} df_3 p(f_3)
\eea
which reduces to the second integral in (\ref{recvalw00})  for $L \gg 1$. Thus for large $L$, the cumulative distribution ${\cal P}_R(D,w)$ for odd $D$ is also a function of $r$. However 
unlike extreme value distribution for odd $D$, the  distributions for even and odd $D$ do not  match for $L \gg 1$ as the expression for the distributions for the distributions ${\cal P}(w_{1,0} \leq w)$ and ${\cal P}(w_{0,0} \leq w)$ do not coincide. 

\subsubsection{Distribution of the number of records}
\label{recn}

To find the probability $N_R(n)$ that the total number of records equals $n$, we first  calculate the record configuration probability $Q(\{w_{n_1,n_2}(D) \})$ defined as the probability that all the elements in the set $\{w_{n_1,n_2}(D) \}$ are records. This distribution depends on the location of the global maximum. 
 If $f_{0}$ is the largest block fitness, the global maximum occurs at $D=0$ and obviously there are no records beyond $D=0$ in this case. 
 
 When $f_0$ is not a global maximum and $f_1 > f_2$, only four record configurations occur with a nonzero  probability. 
 When the fittest block has a fitness $f_{1}$, a record cannot occur beyond $D=L/2$ and  only the conditions in (\ref{recwd0}) are satisfied since  $2 f_1-f_0-f_3$ must be positive. Thus the fitness $w_{D,0}(D)$ for all $D \leq L/2$ is a record with probability 
\be
Q(w_{1,0}(1),...,w_{L/2,0}(L/2))=\frac{1}{4}
\label{spc0}
\ee

When the block fitness $f_{3}$ is the largest, the records occur until $D=L$ at a spacing of one or two  depending on the sign of $f_1 -f_0$ as explained below:

(i) From
the discussion at the beginning of Sec.~\ref{rec}, it is evident that when $f_1 < f_0$,  the only set of fitnesses that can be a record are $w_{0,0}(D)$ for all $D$. Using the conditions in (\ref{recw00}), it follows that when $f_2 < f_1< f_0 < f_3$ , a record occurs only in even $D$ shells. As $f_i$'s are independent and identically distributed (i.i.d.)  random variables, all $4!$ block fitness configurations are equally likely and therefore we get    
\be
Q(w_{0,0}(2),w_{0,0}(4),...,w_{0,0}(L))= \frac{1}{24} \label{spc1}
\ee

(ii) If $f_1 > f_0$ (and $f_2$), the fitness $w_{1,0}(1)$ is a record. The next record depends  on the sign of $2 f_1-f_0-f_3$. From (\ref{recw00}) and (\ref{recw10}), it follows that if $2 f_1- f_0-f_3 < 0$, the fitness $w_{0,0}(D)$ is a record for all even $D$ and $w_{1,0}(D)$ for all odd $D$ with probability 
\be
Q(w_{1,0}(1),w_{0,0}(2),...,w_{1,0}(L-1),w_{0,0}(L))=
 \int_{l} ^{u} df_{0} p(f_{0}) \int_{f_0}
^{u} df_{1} p(f_{1}) \int_{l} ^{f_{1}}df_{2}
p(f_{2}) \int_{2f_{1}-f_{0}} ^{u}df_{3} p(f_{3})
\label{spc2}
\ee 
 If $2 f_1-f_0-f_3> 0$,  due to (\ref{recwd0}) and (\ref{recwld0}),
the fitnesses $w_{D,0}(D)$ for all $D \leq L/2$ and $w_{L-D,0}(D)$ for all $D > L/2$ are records. This event occurs with probability 
\be
Q(w_{1,0}(1),...,w_{L/2,0}(L/2),w_{L/2-1,0}(L/2+1),...,w_{0,0}(L))= \int_l^u df_0 p(f_0) \int_{f_0}^u
df_1 p(f_1) \int_l^{f_1} df_2
p(f_2) \int_{f_1}^{2 f_1-f_0} df_3 p(f_3)
\label{spc3}
\ee

From the above discussion, it is evident that  the total number of records (ignoring the one at $D=0$) can be either $L/2$ (due to (\ref{spc0}) and (\ref{spc1})) or $L$ (see (\ref{spc2}) and (\ref{spc3})). The probability $N_R(n)$ of total number $n$ of records  is independent of underlying block fitness distribution and is given by
\be
N_R(L/2)=2 \left(\frac{1}{4}+\frac{1}{24}\right)=\frac{7}{12}~,~N_R(L)= \frac{2}{12}=\frac{1}{6}
\ee
where we have used that twice the sum of (\ref{spc2}) and (\ref{spc3}) equals (\ref{odd2}).
The average number ${\cal R}$ of records can be found using $N_R(n)$ or $P_R(D)$  and is given by 
\be
{\cal R}=\sum_{n=1}^L n N_R(n)=\sum_{D=1}^L P_R(D)= \frac{11 L}{24} \approx 0.458 L
\label{recn2}
\ee
for any even $L$. 

\subsection{Reaching the global maximum}
\label{jum}

As discussed in Sec.~\ref{models}, all records are contenders for being a leader; however only those records for which the overtaking time is minimised qualifies to be a jump \cite{Krug:2003,Jain:2005,Jain:2007c}. Like records, the statistics of jumps depends on the location of the global maximum.
If $f_{0}$ is the fittest block, the unmutated sequence with fitness
$w_{0,0}(0)=f_0$ is the leader throughout. 

 If $f_1 (> f_2$) is the global maximum,  the last record and hence the last jump occurs at $D=L/2$. Since the time of intersection $T(0,D)$ of the population $E(D,t), D \leq L/2$ with the population $E(0,t)$ given by
\begin{eqnarray}
 T_1=T(0,D)=\frac{D}{w_{D,0}(D)-w_{0,0}(0)}=\frac{L}{2 (f_{1}-f_{0})}  ~,~D \leq L/2
 \label{jmpf1}
\end{eqnarray}
is independent of $D$, all the populations overtake the population of the initial sequence at the
same point. Thus all the record populations participate in the evolutionary race. But as the population  $E(L/2,t)$ has the largest fitness, it becomes the final leader thus leading to a single jump when $f_1$ (or $f_2$) is the largest fitness.

If the global maximum is $f_3$ which occurs at $D=L$,  the following cases  as discussed in Sec~\ref{recn} arise:

(i) If $f_1 < f_0$,  the population with the record fitness $w_{0,0}(D), D \leq L$ overtakes that with the initial fitness $w_{0,0}(0)$ at a time given by
\be
T_{3,1}=T(0,D)=\frac{D}{w_{0,0}(D)-w_{0,0}(0)}=\frac{L}{f_3-f_0} ~,~D \leq L
\label{T31}
\ee
so that all the populations with record fitness $w_{0,0}(D)$ intersect at the same time and the population of the global maximum at $D=L$ takes over in a single jump.

(ii) If $f_1 > f_0$ and $2 f_1-f_0-f_3 < 0$, the population with fitness $w_{0,0}(D)$ for all even $D$  and $w_{1,0}(D)$ for all odd $D$ intersects $E(0,t)$ at the following intersection time:
\begin{eqnarray}
T(0, D)&=& \frac{D}{w_{1,0}(D)-w_{0,0}(0)}=\frac{LD}{(D-1)f_{3}+2f_{1}-(D+1)f_{0}} ~,~\text{for odd}~D\\
T(0,D)&=&\frac{D}{w_{0,0}(D)-w_{0,0}(0)}=\frac{L}{f_{3}-f_{0}}
~,~\text{for even}~D 
\label{T32}
\end {eqnarray}
By virtue of the condition $2 f_1-f_0-f_3 < 0$, the intersection time for odd $D$ is greater than that for even $D$. Therefore the current leader at $D=0$ is overtaken by $D=L$ resulting in a single jump at time $T_{3,2}=L/(f_3-f_0)$. 

If $2 f_1-f_0-f_3 > 0$, the record fitnesses are $w_{D,0}(D)$ for $D \leq L/2$ and $w_{L-D,0}(D)$ for $D > L/2$. The 
populations corresponding to these fitnesses overtake the leader at $D=0$ at time 
\begin{eqnarray}
T(0,D)&=&\frac{L}{2 (f_{1}-f_{0})}~,~D \leq L/2 \\
T(0,D)&=&\frac{D L}{(2D-l)f_{2}+2 (L-D)f_{1}-L f_{0}}~,~D > L/2
\end{eqnarray}
As the intersection time for $D \leq L/2$ is minimum amongst the rest and $w_{L/2,0}(L/2)$ is the largest fitness, the first jump occurs when the population of the sequence with fitness $w_{L/2,0}(L/2)$ overtakes $E(0,t)$.  
 The next change in leader occurs at 
the point of intersection of populations involving the fitness $w_{L-D,0}(D), D>L/2$ with the current leader at a time 
\begin{eqnarray}
T_{3,3}=T(L/2, D)&=&\frac{L}{2 (f_{3}-f_{1})}~,~D > L/2
\label{T33}
\end{eqnarray}
which is again $D$ independent. Thus the population $E(L,t)$ 
is the leader after $E(L/2,t)$ and the global maximum is reached in two jumps.

\subsubsection{Distribution of the number of jumps}

It is obvious that when any block fitness other than $f_0$ is the globally largest fitness, there
will be at least 
one jump (corresponding to globally fittest being the final leader) so that the probability of at least one jump equals $3/4$. In addition, 
there can be one 
more jump when $f_3$ is the global maximum and $2f_{1}-f_{0}-f_{3} > 0$ (see (\ref{T33})).  Due to (\ref{spc3}), the probability $p_2$ of the second jump is given by 
\be
p_2=2 \int_l^u df_0 p(f_0) \int_{f_0}^u
df_1 p(f_1) \int_l^{f_1} df_2
p(f_2) \int_{f_1}^{2 f_1-f_0} df_3 p(f_3) \Theta(u-2f_1+f_0)
\ee
Thus the average number ${\cal J}$ of jumps is given by $(3/4)+p_2$. As $p_2$ is independent of $L$, the average number of jumps is of order unity for any underlying distribution but the constant $p_2$ is not universal. For instance,  when the block fitnesses are chosen from an exponential probability distribution, 
$p_2=5/72 \approx 0.069$ while for uniform distribution, it equals $5/48 \approx 0.104$.

\subsubsection{Temporal jump distribution}
\label{temp}

\begin{figure}
\begin{center}
\includegraphics[angle=270,scale=0.4]{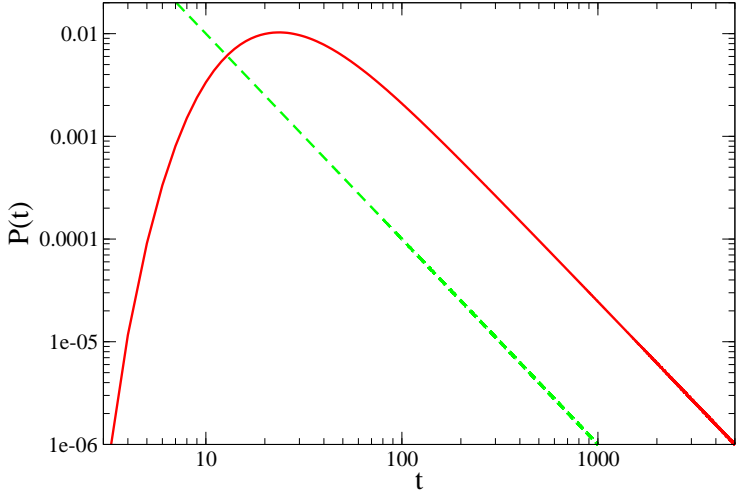}
\caption{(Color online) Log-log plot of the distribution $P(t)$ of the last jump for $p(f)=e^{-f}$  and $L=100$. The broken line has a slope $-2$.}
\label{tempfig}
\end{center}
\end{figure}

We are interested in the probability $P(t)$ that the last jump occurs at time $t > 0$ shown in Fig.~\ref{tempfig} for $p(f)=e^{-f}$. This distribution is a sum of the probability $P_A(t)$ that the last jump occurs at $t$ when $f_1$ or $f_2$ is a global maximum and $P_B(t)$ when $f_3$ is a global maximum.
  We first consider the cumulative probability ${\cal P}_A(t)=\int_0^t dt' P_A(t')$ which on using that $f_1$ (or $f_2$) is a global maximum and (\ref{jmpf1}) gives 
\bea
{\cal P}_A(t) &=& 2  \int_l^{u} \prod_{i=0}^3 df_i
p(f_i) \Theta(t-T_1) \Theta(f_1-f_0) \Theta(f_1-f_2) 
 \Theta(f_1-f_3)  \\
&=& 2 \int_{l+\frac{L}{2 t}}^u df_1 p(f_1) \int_l^{f_1-\frac{L}{2 t}} df_0 p(f_0) \int_l^{f_1} df_2 p(f_2) \int_l^{f_1} df_3 p(f_3)
\eea
 Differentiating ${\cal P}_A(t)$ with respect to time $t$ yields 
\be
P_A(t)=\frac{-L}{2 t^2} \frac{d {\cal P}_A}{d \epsilon}=\frac{L}{t^2} \int_{l+\epsilon}^u df p(f) p(f-\epsilon)  \left( \int_l^f dg p(g) \right)^2
\ee
where we have defined $\epsilon=L/( 2 t)$. For large times $t \gg L/2$,  the integral on the right hand side of the above equation reduces to the probability $G(0)$ that the gap between the globally largest and the second largest in a set of i.i.d. random variables is zero \cite{Jain:2005}. Thus the probability $P_A(t)$ decays as  $\sim L G(0)/t^2$ at large times.

When $f_3$ is the largest fitness (and $f_1 > f_2$), the last jump can occur at times given by (\ref{T31}), (\ref{T32}) and (\ref{T33}).  As $T_{3,1}=T_{3,2}$, the corresponding conditions (discussed in Sec.~\ref{recn}) on the block fitnesses can be combined to give the following cumulative probability
\bea
{\cal P}_1(t) = \int_{l+2 \epsilon}^u df_3 p(f_3) \int_l^{f_3-2 \epsilon} df_0 p(f_0) \int_{l}^{\frac{f_0+f_3}{2}} df_1 p(f_1) \int_l^{f_1} df_2 p(f_2) 
\eea 
and the probability distribution 
\be
P_1(t)=\frac{L}{t^2} \int_{l+2 \epsilon}^u df_3 p(f_3) p(f_3-2 \epsilon)  \int_{l}^{f_3- \epsilon} df_1 p(f_1) \int_l^{f_1} df_2 p(f_2) 
\ee
which also decays as $1/t^2$ at large times. 
 An expression for the distribution for the last jump time $T_{3,3}$ can also be written down in an analogous manner and reads as 
\bea
P_2(t) =\frac{L}{2 t^2} \int_{l+\epsilon}^u df_1 p(f_1) p(f_1+\epsilon) \int_{l}^{f_1-\epsilon} df_0 p(f_0) \int_l^{f_1} df_2 p(f_2) \stackrel{\epsilon \to 0}{\rightarrow} \frac{L}{2 t^2} G(0) 
\eea
Clearly the distribution $P_B(t)=2 (P_1(t)+P_2(t))  \sim t^{-2}$.  Thus the probability distribution $P(t)=P_A(t)+P_B(t)$ obeys the inverse square law for any block fitness distribution. 



\section{Conclusions}
\label{concl}

In this article, we studied a deterministic model \cite{Krug:2003} describing the evolution of a population of self-replicating sequences on a class of strongly correlated fitness landscapes with several fitness peaks \cite{Perelson:1995}.  
 The broad questions addressed in this paper have been studied  on completely uncorrelated fitness landscapes in previous works \cite{Krug:2003,Jain:2005,Jain:2007c}. Here we are interested in finding how the various evolutionary properties are affected when the sequence fitnesses are correlated. 
 
We are primarily interested in the evolutionary dynamics and in particular, the properties of jumps that occur in the population fitness when the most populated sequence changes. As discussed in Sec.~\ref{models}, the largest fitness at a constant Hamming distance from the initial sequence only need to be considered for this purpose. This led us to consider the problem of the extreme statistics of correlated random variables \cite{David:2003,Jain:2009} which has been much less studied than its uncorrelated counterpart. We found that the extreme value distribution is not  of the Gumbel form which is obtained when the  random variables are i.i.d.  and their distribution decays faster than a power law.  In fact, we expect that the universal scaling distributions which depend only on the nature of the tail of the underlying distribution do not exist for such correlated random variables as the number of independent variables namely the block fitnesses is too small.

As the minimum requirement of a sequence to qualify as a leader is that it must be a record, we also studied several record properties of correlated variables. 
Recently the statistics of record events when the number of observations added at each time step increases either deterministically \cite{Krug:2007} or stochastically \cite{Eliazar:2009} have been studied. The records defined in the shell model are an example of the former category as the number of observations changes as ${L \choose D}$ with $D$. It was shown that the probability for a record to occur in a shell with $D$ mutations is not a continuous function unlike the record distributions for independent  random variables \cite{Jain:2005}; however the universality property that the distribution is   independent of block fitness distribution continues to hold. The average number of records  was found to increase linearly with $L$ as in the maximally uncorrelated case but with the prefactor given by $(1-\ln 2) \approx 0.306$ for the latter case which is smaller than in (\ref{recn2}).

\begin{figure}
\includegraphics[angle=270,scale=0.3]{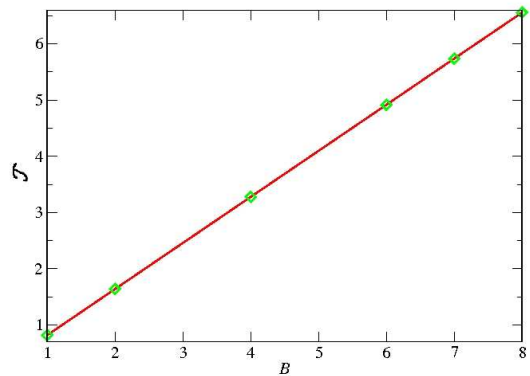}
\caption{(Color online) Average number of jumps as a function of $B$ for the block
  model with locus-dependent block fitness chosen from exponential
  distribution. The line has a slope given by $3/4+p_2=0.819$.}
\label{PereMack}
\end{figure}

 In the uncorrelated fitness model, the  $L$ dependence of the average
 number of jumps was seen to depend on the class of  the  fitness
 distribution $p(f)$. For $p(f)$ decaying faster than a power law, the
 average number of jumps increased as $\sqrt{L}$
 \cite{Krug:2003,Jain:2005}. In contrast, here the average number of
 jumps was shown to be {\it independent} of $L$ for any choice of
 block fitness distribution $p(f)$ although the value of the constant
 was found to be nonuniversal. These results suggest that for block
 fitness distributions decaying faster than a power law, the average
 number of records increases but the average number of jumps decreases
 with increasing correlations. It is also interesting to see how the
 average number of jumps change when the block
 fitness depends not only on the block configuration but also on its
 position in the sequence \cite{Perelson:1995}. The result of our numerical
 simulations for this general model shows that the average number of
 jumps increases linearly with the number of blocks. However the prefactor
 is given by the average number of jumps obtained in the
 locus-independent block fitness model  namely $(3/4)+p_2$ (see
 Fig.~\ref{PereMack}). This suggests that the different blocks
 behave independently in the locus-dependent block fitness model; a detailed
 understanding of this model is beyond the scope of this article and will be  presented elsewhere. 

The temporal distribution for the last jump to occur at time $t$ obeys $t^{-2}$ law for infinite (and finite) populations evolving on uncorrelated fitness landscapes \cite{Krug:2003,Jain:2005,Jain:2007c}. Here we have shown that on a class of strongly correlated fitness landscapes, the same law is obeyed. The origin of this power law can be understood using a simple scaling argument when the fitness variables are independent variables \cite{Krug:2003} but it is not obvious at the outset that such an argument can be used here since the sequence fitnesses are correlated. But it turns out that the jump time involves the i.i.d. block fitnesses and therefore $t^{-2}$  law is obtained here as well. 

We close this article by a discussion of the deterministically evolving populations of infinite size studied here  vis-a-vis 
  finite populations that are subject to stochastic fluctuations on multi-peaked fitness landscapes. 
    As discussed in Sec.~\ref{models}, the basic difference between a finite and an infinite population is that while the former has a finite mutational spread 
in the sequence  space, all the mutants are available at all times in the deterministic case.  As a consequence, on rugged fitness landscapes, a finite population can get trapped at a local optimum from which it can escape by {\it tunneling} through a fitness valley \cite{Jain:2007a}. In fact at late times, most of the population passes exclusively through the local fitness peaks and thus such sequences are the most populated ones when the population size is finite. 
In contrast, as the entire sequence space is occupied for infinite population, a transition to a higher fitness peak takes place by {\it overtaking} the less fitter populations as explained in Sec.~\ref{models}. Thus the underlying mechanism for the punctuated increase of fitness on fitness landscapes with multiple peaks is different in the two situations  \cite{Jain:2007c}. Moreover  the most populated sequence involved in the jump event is not necessarily a local maximum (for any correlation) for infinite populations. 
To see this, consider the fittest sequence with fitness $w^{(max)}(D)$ at Hamming distance $D$ from the initial sequence $\sigma^{(0)}$. Barring the initial sequence, all the one-mutant neighbors of sequence with fitness $w^{(max)}(1)$ are at Hamming distance two from the initial sequence. Consider the scenario when the sequence with fitness $w^{(max)}(2)$ is a nearest neighbor of sequence with fitness $w^{(max)}(1)$. Then the fittest sequence at distance unity from the initial sequence can be a jump if at least $w^{(max)}(1) > w(\sigma^{(0)})$ and the minimum intersection time condition 
$(w^{(max)}(1)-w(\sigma^{(0)}))^{-1} < 2 (w^{(max)}(2)-w(\sigma^{(0)}))^{-1}$ is obeyed. Clearly the latter condition rewritten as $w^{(max)}(2) -w^{(max)}(1) < w^{(max)}(1)-w(\sigma^{(0)})$ can be satisfied even when $w^{(max)}(1)$ is not a local maximum. Thus the number of jump events are not related to the number of local optima for an infinite population. 

Acknowledgments: The authors thank Gayatri Das, Joachim Krug and
Sanjib Sabhapandit for useful discussions.

%

\end{document}